\documentclass[inline]{nature}

\usepackage{amsmath}
\usepackage{amssymb}
\usepackage{latexsym}

\ifx\pdfoutput\undefined
\usepackage{graphicx}
\else
\usepackage[pdftex]{graphicx}
\usepackage{epstopdf}
\fi

\title{Macroscopic effects of the spectral structure in turbulent flows}

\author{Tuan Tran$^1$,  Pinaki Chakraborty$^2$, Nicholas Guttenberg$^3$, 
Alisia Prescott$^4$, Hamid Kellay$^5$,  Walter Goldburg$^4$,
  Nigel Goldenfeld$^{3\star}$ and Gustavo Gioia$^{1\star}$
}

\begin{document}
\def\Re{\textrm{Re}}

\maketitle

\spacing{1}\begin{affiliations}
\item Department of Mechanical Science and Engineering, University of
Illinois, Urbana, IL 61801, USA
\item Department of Geology, University of Illinois, Urbana, IL 61801, USA
\item Department of Physics, University of Illinois,
 Urbana, IL 61801, USA
\item Department of Physics and Astronomy, University of Pittsburgh,
Pittsburgh, PA 15260, USA
\item  {Centre de Physique Mol\'eculaire Optique et Hertzienne,
  Universit\'e Bordeaux I, 
 33405 Talence, France \\ $^\star$  These authors contributed equally to this work } 
\end{affiliations}

\spacing{1}\begin{abstract}

  Two aspects of turbulent flows\cite{FRIS95,SREE99,pope2000tf,hof2006flt}
 have been the subject
 of extensive, split research efforts:
  macroscopic properties, such as the 
  frictional drag\cite{schlichting2000blt}
  experienced by a flow past a wall,
  and the turbulent spectrum.\cite{taylor1938spectrum,FRIS95,SREE97} 
  The turbulent spectrum may be said to represent 
  the fabric of a turbulent state; 
  in practice it is a power law of 
 exponent $\boldsymbol{ \alpha}$ (the ``spectral exponent'') that 
 gives the revolving velocity of a
  turbulent fluctuation (or ``eddy'') of 
 size $\boldsymbol{s}$ as
 a function of $\boldsymbol{s}$.\cite{FRIS95}  
 The link, if any,
 between macroscopic properties and the turbulent spectrum
 remains missing. Might it be found by contrasting
 the frictional drag in
 flows with differing types of spectra?
  Here we perform unprecedented measurements of the
frictional drag in soap-film flows,\cite{KELL02} where 
 the spectral exponent
  $\boldsymbol{\alpha=3}$,\cite{KRAI67,batchelor1982tht}
 and compare the results with the frictional drag 
 in pipe flows,\cite{schlichting2000blt} 
 where the spectral exponent 
 $\boldsymbol{\alpha=5/3}$.\cite{richardson26,KOLM41}
   For moderate values of the Reynolds 
 number $\Re$ (a measure of the strength of the turbulence),
 we find that in soap-film flows
  the frictional drag scales as $\boldsymbol{\Re^{-1/2}}$, whereas 
  in pipe flows the frictional drag scales\cite{mckeon2005ffs}
  as $\boldsymbol{\Re^{-1/4}}$. Each of these scalings may be 
 predicted from the attendant 
 value of 
 $\boldsymbol{\alpha}$ by 
using a new theory,\cite{GIOI06, GUTT09} 
in which the frictional drag is explicitly linked
 to the turbulent spectrum.
Our work indicates that in turbulence, 
as in continuous phase transitions,
 macroscopic properties are governed by the spectral structure of
 the fluctuations.\cite{GOLD06,MEHR08}

\end{abstract}

Turbulent flows past a wall experience frictional 
 drag, the macroscopic property of a flow that sets the cost 
of pumping oil through a pipeline,
the draining capacity of a river in flood,
and other quantities of engineering
 interest\cite{NIKU33,SREE99,pope2000tf,schlichting2000blt,JIME04}.
The frictional drag is 
defined as the dimensionless ratio $f=\tau/\rho U^2$,
 where $\tau$ is the shear stress or force per unit area that
develops between the flow and the wall,
 $\rho$ is the density of the fluid,
 and $U$ is the mean velocity of the flow. 
Already in XVIII Century France, $f$
  was the subject of large-scale experiments
 performed in connection with the design of 
 a waterworks for the city of Paris.\cite{CHEZY1775, DOOGE92} 
Modern experiments with  pipes 
have shown that $f$ depends on the Reynolds
number ${\rm Re}= U d/\nu$, 
where $d$ is the diameter of the pipe
and $\nu$ is the kinematic viscosity of the fluid.
For pipe flows of moderate turbulent strength 
 ($\rm  Re$ up to 98,000) 
the experimental results are well described
(within 1.4\% error\cite{mckeon2005ffs})
 by the Blasius empirical scaling,
  $f \propto {\rm Re}^{-1/4}$. 
 (Throughout this paper, the symbol ``$\propto$''
 may be changed to the symbol ``$=$'' by 
  introducing a dimensionless proportionality factor,
 e.g., $f=C\, {\rm Re}^{-1/4}$.)
A celebrated theory\cite{schlichting2000blt,pope2000tf}
 of the frictional drag 
  was formulated
eighty years ago by Ludwig Prandtl, the founder of 
 turbulent hydraulics, and
 numerous variants\cite{schlichting2000blt,mckeon2005ffs,ALLE07}
 and alternatives\cite{BARE93,BARE04} of Prandtl's theory have since
 been proposed.
Although Prandtl's theory and its variants and alternatives
 yield disparate mathematical
expressions for $f$ as a function of $\Re$,
  for moderate values of $\Re$ they all
give predictions in good numerical accord with the
Blasius empirical scaling.
Yet these theories have been predicated on
dimensional analysis and similarity assumptions,
without reference to the spectral
  structure of the turbulent fluctuations.
As a result, these theories cannot be used to 
reveal the missing link between the frictional drag and 
the turbulent spectrum.


The turbulent spectrum is a function of the wavenumber $k$, $E(k)$,
 whose physical significance may be grasped from the expression 
 $u_s\propto (\int_{1/s}^\infty E(k) dk)^{1/2}$, which gives the 
 revolving velocity $u_s$ of a turbulent eddy 
 of size $s$ in the flow. In general, we can write\cite{GUTT09} 
 $E(k) \propto U^2 L^{(1-\alpha)} k^{-\alpha}$, and therefore 
\begin{equation} \label{theq}
u_s\propto U (s/L)^{(\alpha-1)/2},
\end{equation} 
 where $\alpha$ is the spectral exponent,
  $U$ is the mean velocity
 of the flow, and $L$ is a characteristic length. 
 A single  type of spectrum is possible 
  in three-dimensional (3D) flows: 
 the ``energy cascade,'' for 
 which $\alpha=5/3$.\cite{KOLM41,OBUK41} 
  Thus, e.g., in turbulent pipe flows
 the spectral exponent is $5/3$ and $L=d$, the diameter
 of the pipe. Two-dimensional (2D) turbulence 
 (a type of turbulence that may be realized in a soap film)
 differs from 3D turbulence in a
 number of crucial respects, most notably 
 in that in two dimensions there is no vortex
  stretching.
As a result, a different type of 
spectrum is possible in 2D flows: the ``enstrophy cascade,'' for which 
 $\alpha=3$.\cite{batchelor1982tht,KRAI67} Thus, e.g., in
  turbulent soap-film flows the spectral exponent is $3$ and 
 $L=w$, the width of the soap film. 

\begin{figure}[hpdp]
\begin{center}
\includegraphics[width=9cm]{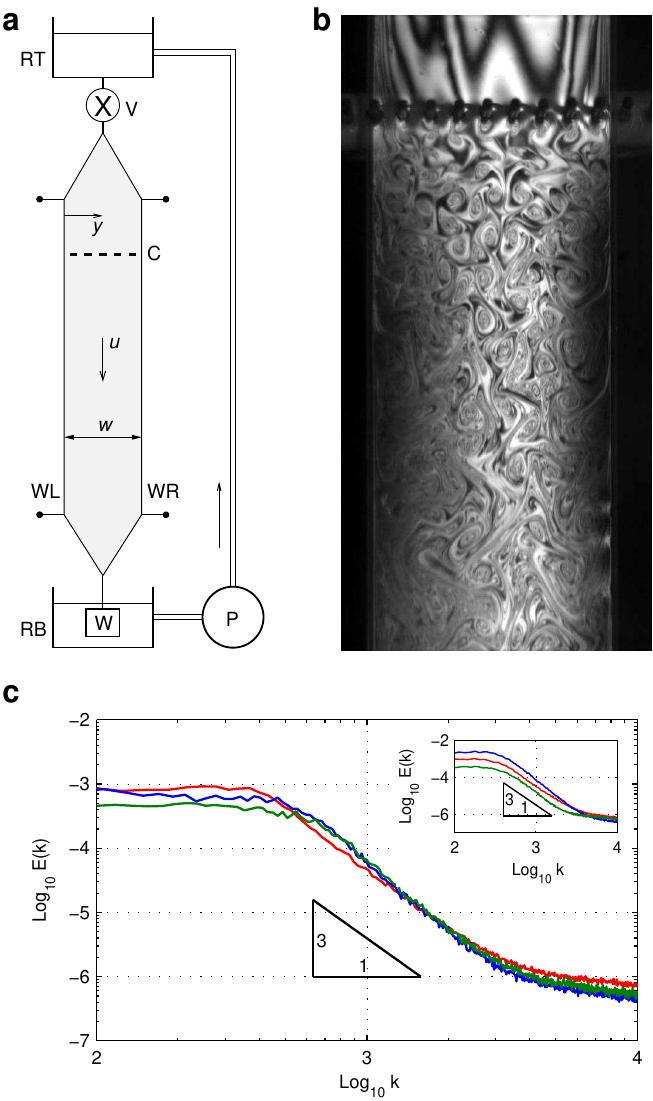}
\spacing{1} \caption{\small{Experimental setup used to study 
   steady, gravity-driven, 2D soap-film flows. {\bf (a)} 
 Wires WL and WR are thin nylon wires
 (diameter $=0.5\,$mm) kept taut by  
 weight W. The film hangs from the wires;
 its width increases from 0 to $w$ over
 an expansion section,
  then remains constant and equal to $w$ over
 a section of length $\approx 1\,$m.
 Reservoir RT contains a soapy solution 
 ($2.5\,$\% Dawn Nonultra in water; $\nu=0.01\,$cm$^2$/s),
 which flows through valve V and into the film.
 After flowing through the film, 
 the soapy solution drains into reservoir RB 
  and returns to reservoir RT via pump P.
  Turbulence is generated by comb C
 of tooth diameter $\approx 0.5\,$mm and
  tooth spacing $\approx 3\,$mm.
 We perform all measurements at a distance 
 of at least 10$\,$cm downstream of the comb.
 Axis $y$ ($0\le y\le w$) has its origin at wire WL, 
 as indicated.
 {\bf (b)} Interference 
 fringes in yellow light (wavelength $=0.6\,\mu$m)
 make it possible to visualize the generation of 2D turbulence
from the comb. {\bf (c)}  Typical log-log plots 
 of the spectrum $E(k)$, from LDV measurements 
 performed on points of the film 
 close to one of the wires and (inset)
  along the centerline of the flow
  }} \label{setup}
\end{center}
\end{figure}

 To study soap-film flows we hang a soap film 
 between two long, vertical, mutually parallel
  wires a few centimeters apart from one another
   (Fig.~\ref{setup}a). 
  Driven by gravity, a steady vertical flow soon 
 becomes established within the film.
 Then, the thickness $h$ of the film is roughly
 uniform on any 
 cross section of the film, 
 typically  $h\approx 10\,\mu$m, much smaller
 than the width $w$  and the length 
 of the film  (Fig.~\ref{setup}a). As a result,
 the velocity of the flow 
 lies on the plane of the film, and 
 the flow is 2D. 

We make the flow turbulent by piercing
 the film with a comb, as indicated in 
 Fig.~\ref{setup}a, so that the flow is stirred as it
 moves past the teeth of the comb.
To visualize the flow, 
 we cast monochromatic light
 on a face of a film and observe the 
 interference fringes that form there. 
 These fringes (Fig.~\ref{setup}b) 
 reflect small changes 
 in the local thickness of the film. 
(The thickness is constant along
 a fringe; it differs by one-half
  wavelength of 
 the light, or a fraction of a $\mu$m,
 between any two successive fringes.)
The small changes in thickness in turn reflect 
 small changes in the absolute value of the 
 instantaneous velocity of the flow.\cite{KELL02}    
Thus Fig.~\ref{setup}b
 may be interpreted 
 as a map of the instantaneous spatial distribution 
 of turbulent fluctuations downstream of the comb.

 We compute the  spectrum $E(k)$ at numerous
  points  on the film from measurements performed 
 with a Laser Doppler Velocimeter (LDV; see Methods).  
 In Fig.~\ref{setup}c we show 
 a few typical log-log plots of $E$ vs.\ $k$.
 The slope of these plots 
 represents the spectral exponent $\alpha$; 
 in our experiments the slope 
 is slightly larger than 3, consistent with
  prior experiments with
 soap-film flows \cite{kellay1995experiments,KELL02},
 and close to the theoretical value of $\alpha$ 
 for the enstrophy cascade ($\alpha=3$). 

\begin{figure}[hpdp]
\begin{center}
\includegraphics[width=9cm]{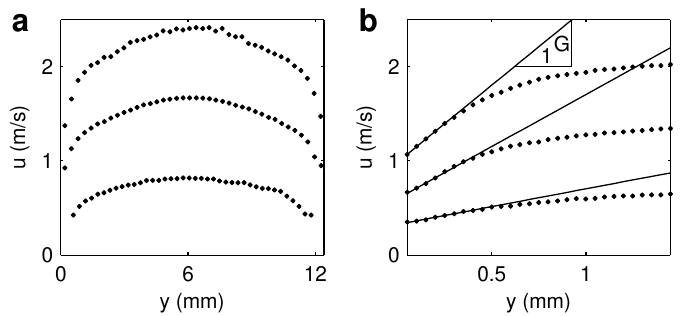}
\spacing{1}
\caption{\small{ The mean velocity profile $u(y)$ in 
turbulent soap-film flows, from LDV measurements.
 {\bf (a)} Typical plots of $u(y)$ in a film of width 
 $w=12\,$mm. {\bf (b)} Typical 
  plots of $u(y)$ close to one of the wires,
  in the viscous layer where $du(y)/dy=G$ and the thickness 
 of the film is uniform and $\approx 10\,\mu$m
  (Supplementary Information). 
The velocity profiles correspond to $\Re = 7893, 17648, 25912$.
Points on the film
 closer than $\approx 20\mu$m (the diameter 
  of the beam of the LDV) from the edge of the wire cannot
 be probed with the LDV; thus the first data point, 
 which we position at $y=0$, is at a distance of 
 $\approx 20\mu$m from the edge of the wire.
 The apparent slip velocity is likely to 
 represent 3D and surface-tension effects associated with 
 the complex flow at the contact 
 between the film and the wire.
}}
\label{velpro}
\end{center}
\end{figure}

 By using the same Laser Doppler Velocimeter
  we measure the mean (time-averaged) 
 velocity $u$ at any point on the film (see Methods).
 Successive measurements of $u$ along a
 cross section of the film 
 gives the ``mean velocity profile''
  $u(y)$ of that cross section.
   In Fig.~\ref{velpro}a we show a few typical
 plots of $u(y)$ over the entire width of the film
  (i.e., from wire to wire, or for $0\le y\le w$;
  see Fig.~\ref{setup}a).
  From a mean velocity profile
  we compute the mean velocity of the flow as
 $U=(1/w)\,\int_0^w u(y)dy$, and the 
 Reynolds number as ${\rm Re}=U w/\nu$.
 
\begin{figure}[hpdp]
\begin{center}
\includegraphics[width=9cm]{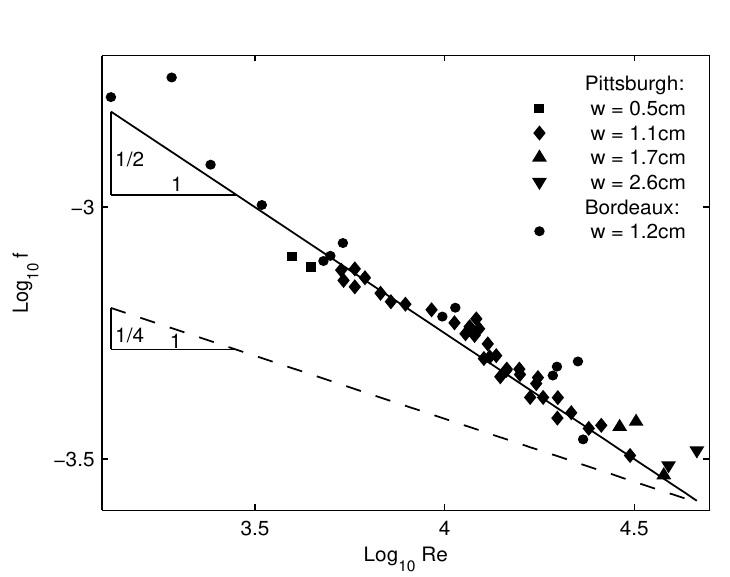}
\spacing{1}
\caption{\small{ 
Log-log plot of the frictional drag vs.\ the
 Reynolds number in 2D
 turbulent soap-film flows of Reynolds number
  $1300\le{\rm Re}\le 25000$, from independent
 experiments performed in Pittsburgh and Bordeaux.
   The cloud of data points may be represented as
  a straight line of slope $1/2$, consistent 
 with the scaling $f \propto {\rm Re}^{-1/2}$. The straight 
 dashed line of slope $1/4$ corresponds to the 
 Blasius empirical scaling, $f \propto {\rm Re}^{-1/4}$. }}
\label{friction_factor}
\end{center}
\end{figure}

 In Fig.~\ref{velpro}b we show a few typical 
 plots of $u(y)$ close to one of the wires,
 where the mean velocity profile is linear
 on a narrow  (about 0.2$\,$mm) viscous layer.
 From the slope $G$ 
 of the mean velocity profile in the viscous layer, 
  we compute the shear stress between the flow
 and the wire as $\tau=\rho \nu G$, and 
 the frictional drag as $f=\tau/\rho U^2=\nu G/U^2$.
 In Fig.~\ref{friction_factor} we show
 a log-log plot of $f$ vs.\ Re.
 The plot consists of five sets
of data points from numerous turbulent soap-film flows; 
 four sets were taken at Pittsburgh, and
one at Bordeaux in an independent experimental setup.
The cloud of data points is consistent 
 with the scaling, $f\propto {\rm Re}^{-1/2}$,
 and inconsistent with the Blasius empirical scaling, 
 $f\propto {\rm Re}^{-1/4}$, which is known
 to prevail 
  in turbulent pipe flows.

Our experimental results may be explained 
 using a recently proposed
  theory of the frictional drag.\cite{GIOI06,GUTT09} 
  In this new theory, the frictional drag 
 is produced by turbulent eddies that transfer momentum
  between the wall or wire (where the fluid carries a negligible 
 momentum per unit mass) and the turbulent flow (where the
 fluid carries a sizable momentum 
 per unit mass). The theory is generally applicable
 to flows that move past rough walls, but for 
 the special case in which the wall is smooth
  (as in our experiments), the theory predicts that
  $f\propto u_\eta/U$,\cite{GIOI06,GUTT09,GB02,GCB06,calzetta2009friction,MEHR08}  where $\eta$ is the size of the smallest eddies in the flow and
 $u_\eta$ is the revolving velocity of those eddies.
 Note that in this scaling for $f$,
  $u_\eta$ provides a connection to the 
 turbulent spectrum, as shown next.


To a generic eddy of size  $s$ 
  we ascribe an eddy Reynolds number ${\rm Re}_s=u_s s/\nu$,
 where $u_s$ is the revolving velocity of the eddy.
 The Reynolds number of the smallest eddies in the flow
 is of order 1 (because these eddies 
 are viscous\cite{KOLM41,FRIS95}), and we can write
  ${\rm Re}_\eta =u_\eta \eta/\nu\approx 1$, or 
 $u_\eta\propto \nu/\eta$.
 A second expression for $u_\eta$ may be obtained 
 from the spectrum by 
 setting $s=\eta$ in (\ref{theq}); if we combine
  these two expressions so as to 
 eliminate $\eta$,   we can 
 write $u_\eta \propto U\, {\rm Re}^{(1-\alpha)/(1+\alpha)}$,
 where we have used the definition ${\rm Re}=U L/\nu$.
 As $f\propto u_\eta/U$, it follows that 
 \begin{equation} \label{fofalpha}
f\propto {\rm Re}^{(1-\alpha)/(1+\alpha)},
\end{equation}
 and the functional dependence 
 of the frictional drag on the Reynolds number 
 is set by the spectral exponent $\alpha$.
 For pipe 
 flows $\alpha=5/3$, and (\ref{fofalpha}) yields 
 the prediction\cite{GIOI06}
 $f\propto {\rm Re}^{-1/4}$, consistent
  with the Blasius scaling. In contrast, 
for soap-film flows $\alpha=3$, and 
 (\ref{fofalpha}) yields the prediction\cite{GUTT09} 
 $f\propto {\rm Re}^{-1/2}$, consistent
  with our experimental results 
 (Fig.~\ref{friction_factor}).

 From our experiments with two-dimensional soap-film flows
 we infer that the long-standing and
widely-accepted theory\cite{schlichting2000blt}
  of the frictional drag between a turbulent
flow and a wall is incomplete. This classical
theory does not take into account the structure of the 
 turbulent fluctuations, and cannot distinguish 
 between two-dimensional and three-dimensional 
 turbulent flows. Our  data on soap-film flows,
 as well as the available data on  
 pipe flows, are, however, consistent with the predictions 
of a recently proposed theory of the frictional
  drag.\cite{GIOI06,GUTT09} This new theory
  perforce relates the frictional drag 
  to the turbulent spectrum, and is sensitive 
 to the dimensionality of the flow via the dependence
 of the turbulent spectrum on the dimensionality.
Our findings lead us to conclude
  that the macroscopic properties of both three-dimensional
 and two-dimensional turbulent flows are linked to 
 the turbulent fluctuations, in a manner analogous to
 the way in which order-parameter fluctuations are linked to
 the macroscopic properties of thermodynamic
  systems close to a critical point.\cite{GOLD06,MEHR08}
 In addition, our findings serve to underscore the value of using
 two-dimensional soap-film
 flows to test and extend our understanding of
 turbulent phenomena. 


\begin{methods}
We measure the vertical component $u(t)$ of the 
 instantaneous velocity at a point on the film
 using a Laser Doppler Velocimeter with
 a sampling rate of 5 kHz.
 By performing measurements over a time period 
  of about 10$\,$s, we collect a time series $u(t_i)$.
 From the time series we  
 compute the local mean 
 velocity $u$ as the time average 
 of $u(t_i)$, $u=\langle u(t_i)\rangle$. 
 From the same 
 time series $u(t_i)$ we compute the local
 turbulent spectrum (more precisely, the longitudinal 
 turbulent spectrum). To that end, we invoke 
 Taylor's frozen-turbulence hypothesis\cite{taylor1938spectrum}
 to perform a space-for-time substitution
 $t\rightarrow x/u$ on the time series
   $(u(t_i)-u)$ to obtain a space series
  $v(x_i)=(u(x_i/u)-u)$, where $x_i=u t_i$. 
  (The frozen-turbulence 
 hypothesis is justified because in all our experiments the root 
 mean square of the velocity fluctuations is less than 20\% 
 of  $u$.\cite{belmonte2000experimental}) 
The spectrum $E(k)$ is the square of the magnitude of the
 discrete  Fourier transform of $v(x_i)$. 
\end{methods}

\bibliography{friction2d.bib}

\begin{thebibliography}{10}
\expandafter\ifx\csname url\endcsname\relax
  \def\url#1{\texttt{#1}}\fi
\expandafter\ifx\csname urlprefix\endcsname\relax\def\urlprefix{URL }\fi
\providecommand{\bibinfo}[2]{#2}
\providecommand{\eprint}[2][]{\url{#2}}

\bibitem{FRIS95}
\bibinfo{author}{Frisch, U.}
\newblock \emph{\bibinfo{title}{Turbulence: The Legacy of A.N. Kolmogorov}}
  (\bibinfo{publisher}{Cambridge University Press, Cambridge, UK},
  \bibinfo{year}{1995}).

\bibitem{SREE99}
\bibinfo{author}{Sreenivasan, K.~R.}
\newblock \bibinfo{title}{Fluid turbulence}.
\newblock \emph{\bibinfo{journal}{Rev. Mod. Phys.}}
  \textbf{\bibinfo{volume}{71}}, \bibinfo{pages}{S383--S395}
  (\bibinfo{year}{1999}).

\bibitem{pope2000tf}
\bibinfo{author}{Pope, S.}
\newblock \emph{\bibinfo{title}{{Turbulent Flows}}}
  (\bibinfo{publisher}{Cambridge University Press, Cambridge, UK},
  \bibinfo{year}{2000}).

\bibitem{hof2006flt}
\bibinfo{author}{Hof, B.}, \bibinfo{author}{Westerweel, J.},
  \bibinfo{author}{Schneider, T.} \& \bibinfo{author}{Eckhardt, B.}
\newblock \bibinfo{title}{{Finite lifetime of turbulence in shear flows}}.
\newblock \emph{\bibinfo{journal}{Nature}} \textbf{\bibinfo{volume}{443}},
  \bibinfo{pages}{59--62} (\bibinfo{year}{2006}).

\bibitem{schlichting2000blt}
\bibinfo{author}{Schlichting, H.} \& \bibinfo{author}{Gersten, K.}
\newblock \emph{\bibinfo{title}{{Boundary-Layer Theory}}}
  (\bibinfo{publisher}{Springer, New York, USA}, \bibinfo{year}{2000}).

\bibitem{taylor1938spectrum}
\bibinfo{author}{Taylor, G.}
\newblock \bibinfo{title}{{The spectrum of turbulence}}.
\newblock \emph{\bibinfo{journal}{Proceedings of the Royal Society of London.
  Series A, Mathematical and Physical Sciences}} \bibinfo{pages}{476--490}
  (\bibinfo{year}{1938}).

\bibitem{SREE97}
\bibinfo{author}{Sreenivasan, K.~R.}
\newblock \bibinfo{title}{The phenomenology of small-scale turbulence}.
\newblock \emph{\bibinfo{journal}{Annu. Rev. Fluid Mech.}}
  \textbf{\bibinfo{volume}{29}}, \bibinfo{pages}{435--472}
  (\bibinfo{year}{1997}).

\bibitem{KELL02}
\bibinfo{author}{Kellay, H.} \& \bibinfo{author}{Goldburg, W.~I.}
\newblock \bibinfo{title}{Two-dimensional turbulence: a review of some recent
  experiments}.
\newblock \emph{\bibinfo{journal}{Rep. Prog. Phys.}}
  \textbf{\bibinfo{volume}{65}}, \bibinfo{pages}{845--894}
  (\bibinfo{year}{2002}).

\bibitem{KRAI67}
\bibinfo{author}{Kraichnan, R.~H.}
\newblock \bibinfo{title}{Inertial ranges in two-dimensional turbulence}.
\newblock \emph{\bibinfo{journal}{The Physics of Fluids}}
  \textbf{\bibinfo{volume}{10}}, \bibinfo{pages}{1417--1423}
  (\bibinfo{year}{1967}).

\bibitem{batchelor1982tht}
\bibinfo{author}{Batchelor, G.}
\newblock \emph{\bibinfo{title}{{The Theory of Homogeneous Turbulence}}}
  (\bibinfo{publisher}{Cambridge University Press, Cambridge, UK},
  \bibinfo{year}{1953}).

\bibitem{richardson26}
\bibinfo{author}{Richardson, L.~F.}
\newblock \bibinfo{title}{Atmospheric diffusion shown on a distance--neighbour
  graph}.
\newblock \emph{\bibinfo{journal}{Proc. Roy. Soc. London A}}
  \textbf{\bibinfo{volume}{110}}, \bibinfo{pages}{709--737}
  (\bibinfo{year}{1926}).

\bibitem{KOLM41}
\bibinfo{author}{Kolmogorov, A.~N.}
\newblock \bibinfo{title}{Local structure of turbulence in incompressible fluid
  at a very high reynolds number}.
\newblock \emph{\bibinfo{journal}{Dokl. Akad. Nauk. SSSR}}
  \textbf{\bibinfo{volume}{30}}, \bibinfo{pages}{299--302}
  (\bibinfo{year}{1941}).
\newblock \bibinfo{note}{[English translation in Proc. R. Soc. London Ser. A
  434 (1991)]}.

\bibitem{mckeon2005ffs}
\bibinfo{author}{Mckeon, B.~J.}, \bibinfo{author}{Zagarola, M.~V.} \&
  \bibinfo{author}{Smits, A.~J.}
\newblock \bibinfo{title}{A new friction factor relationship for fully
  developed pipe flow}.
\newblock \emph{\bibinfo{journal}{Journal of Fluid Mechanics}}
  \textbf{\bibinfo{volume}{538}}, \bibinfo{pages}{429--443}
  (\bibinfo{year}{2005}).

\bibitem{GIOI06}
\bibinfo{author}{Gioia, G.} \& \bibinfo{author}{Chakraborty, P.}
\newblock \bibinfo{title}{Turbulent friction in rough pipes and the energy
  spectrum of the phenomenological theory}.
\newblock \emph{\bibinfo{journal}{Phys. Rev. Lett}}
  \textbf{\bibinfo{volume}{96}}, \bibinfo{pages}{044502}
  (\bibinfo{year}{2006}).

\bibitem{GUTT09}
\bibinfo{author}{Guttenberg, N.} \& \bibinfo{author}{Goldenfeld, N.}
\newblock \bibinfo{title}{Friction factor of two-dimensional rough-boundary
  turbulent soap film flows}.
\newblock \emph{\bibinfo{journal}{Physical Review E (Statistical, Nonlinear,
  and Soft Matter Physics)}} \textbf{\bibinfo{volume}{79}},
  \bibinfo{pages}{065306} (\bibinfo{year}{2009}).

\bibitem{GOLD06}
\bibinfo{author}{{Goldenfeld}, N.}
\newblock \bibinfo{title}{{Roughness-Induced Critical Phenomena in a Turbulent
  Flow}}.
\newblock \emph{\bibinfo{journal}{Phys. Rev. Lett.}}
  \textbf{\bibinfo{volume}{96}}, \bibinfo{pages}{044503}
  (\bibinfo{year}{2006}).

\bibitem{MEHR08}
\bibinfo{author}{Mehrafarin, M.} \& \bibinfo{author}{Pourtolami, N.}
\newblock \bibinfo{title}{Intermittency and rough-pipe turbulence}.
\newblock \emph{\bibinfo{journal}{Phys. Rev. E}} \textbf{\bibinfo{volume}{77}},
  \bibinfo{pages}{055304} (\bibinfo{year}{2008}).

\bibitem{NIKU33}
\bibinfo{author}{Nikuradze, J.}
\newblock \bibinfo{title}{{Stromungsgesetze in rauhen Rohren}}.
\newblock \emph{\bibinfo{journal}{VDI Forschungsheft}}
  \textbf{\bibinfo{volume}{361}} (\bibinfo{year}{1933}).
\newblock \bibinfo{note}{[English translation available as National Advisory
  Committee for Aeronautics, Tech. Memo. 1292 (1950). Online at:
  http://hdl.handle.net/2060/19930093938]}.

\bibitem{JIME04}
\bibinfo{author}{Jimenez, J.}
\newblock \bibinfo{title}{{Turbulent flows over rough walls}}.
\newblock \emph{\bibinfo{journal}{Annual Review of Fluid Mechanics}}
  \textbf{\bibinfo{volume}{36}}, \bibinfo{pages}{173--196}
  (\bibinfo{year}{2004}).

\bibitem{CHEZY1775}
\bibinfo{author}{Ch\'ezy, A.}
\newblock \bibinfo{title}{Memoire sur la vitesse de l'eau conduit dans une
  rigole donne}.
\newblock In \emph{\bibinfo{booktitle}{Dossier 847 (MS 1915)}}
  (\bibinfo{publisher}{Ecole des Ponts et Chaussees}, \bibinfo{year}{1775}).
\newblock \bibinfo{note}{[English translation in Journal, Association of
  Engineering Societies, vol 18, 363--368, 1897]}.

\bibitem{DOOGE92}
\bibinfo{author}{Dooge, J. C.~I.}
\newblock \bibinfo{title}{The manning formula in context}.
\newblock In \bibinfo{editor}{Yen, B.~C.} (ed.)
  \emph{\bibinfo{booktitle}{Channel flow resistance: centennial of Manning's
  formula}}, \bibinfo{pages}{136--185} (\bibinfo{publisher}{Water Resoures
  Publications, Littleton, Colorado}, \bibinfo{year}{1992}).

\bibitem{ALLE07}
\bibinfo{author}{Allen, J.~J.}, \bibinfo{author}{Shockling, M.~A.},
  \bibinfo{author}{Kunkel, G.~J.} \& \bibinfo{author}{Smits, A.~J.}
\newblock \bibinfo{title}{Turbulent flow in smooth and rough pipes}.
\newblock \emph{\bibinfo{journal}{Phil. Trans. Roy. Soc. A}}
  \textbf{\bibinfo{volume}{365}}, \bibinfo{pages}{699--714}
  (\bibinfo{year}{2007}).

\bibitem{BARE93}
\bibinfo{author}{Barenblatt, G.}
\newblock \bibinfo{title}{{Scaling laws for fully developed turbulent shear
  flows. Part 1. Basic hypotheses and analysis}}.
\newblock \emph{\bibinfo{journal}{Journal of Fluid Mechanics}}
  \textbf{\bibinfo{volume}{248}}, \bibinfo{pages}{513--520}
  (\bibinfo{year}{1993}).

\bibitem{BARE04}
\bibinfo{author}{Barenblatt, G.} \& \bibinfo{author}{Chorin, A.}
\newblock \bibinfo{title}{{A mathematical model for the scaling of
  turbulence}}.
\newblock \emph{\bibinfo{journal}{Proceedings of the National Academy of
  Sciences}} \textbf{\bibinfo{volume}{101}}, \bibinfo{pages}{15023--15026}
  (\bibinfo{year}{2004}).

\bibitem{OBUK41}
\bibinfo{author}{Obukhov, A.~M.}
\newblock \bibinfo{title}{Energy distribution in the spectrum of turbulent
  flow}.
\newblock \emph{\bibinfo{journal}{Dokl. Akad. Nauk. SSSR}}
  \textbf{\bibinfo{volume}{32}}, \bibinfo{pages}{22--24}
  (\bibinfo{year}{1941}).

\bibitem{kellay1995experiments}
\bibinfo{author}{Kellay, H.}, \bibinfo{author}{Wu, X.} \&
  \bibinfo{author}{Goldburg, W.}
\newblock \bibinfo{title}{{Experiments with Turbulent Soap Films}}.
\newblock \emph{\bibinfo{journal}{Physical Review Letters}}
  \textbf{\bibinfo{volume}{74}}, \bibinfo{pages}{3975--3978}
  (\bibinfo{year}{1995}).

\bibitem{GB02}
\bibinfo{author}{Gioia, G.} \& \bibinfo{author}{Bombardelli, F.~A.}
\newblock \bibinfo{title}{{Scaling and similarity in rough channel flows}}.
\newblock \emph{\bibinfo{journal}{Phys. Rev. Lett.}}
  \textbf{\bibinfo{volume}{88}}, \bibinfo{pages}{014501}
  (\bibinfo{year}{2002}).

\bibitem{GCB06}
\bibinfo{author}{Gioia, G.}, \bibinfo{author}{Chakraborty, P.} \&
  \bibinfo{author}{Bombardelli, F.~A.}
\newblock \bibinfo{title}{{Rough-pipe flows and the existence of fully
  developed turbulence}}.
\newblock \emph{\bibinfo{journal}{Phys. Fluids}} \textbf{\bibinfo{volume}{18}},
  \bibinfo{pages}{038107} (\bibinfo{year}{2006}).

\bibitem{calzetta2009friction}
\bibinfo{author}{Calzetta, E.}
\newblock \bibinfo{title}{{Friction factor for turbulent flow in rough pipes
  from Heisenberg's closure hypothesis}}.
\newblock \emph{\bibinfo{journal}{Physical Review E}}
  \textbf{\bibinfo{volume}{79}}, \bibinfo{pages}{56311} (\bibinfo{year}{2009}).

\bibitem{belmonte2000experimental}
\bibinfo{author}{Belmonte, A.}, \bibinfo{author}{Martin, B.} \&
  \bibinfo{author}{Goldburg, W.}
\newblock \bibinfo{title}{{Experimental study of Taylor's hypothesis in a
  turbulent soap film}}.
\newblock \emph{\bibinfo{journal}{Physics of Fluids}}
  \textbf{\bibinfo{volume}{12}}, \bibinfo{pages}{835--845}
  (\bibinfo{year}{2000}).

\end{thebibliography}

\begin{addendum}
\item[Author Information] Reprints and permissions information is available
at npg.nature.com/reprints and permissions. The authors declare that they
have no competing financial interests. 
\item We thank discussions with J.\ M.\ Larkin.  This work was 
 funded by the US National Science Foundation through  
 NSF/DMR grant 06-04477 and NSF/DMR grant 06-04435 
 (W.\ Fuller--Mora, Programme Director).
 T.\ Tuan acknowledges support from the Vietnam Education Foundation.
\item[Author Contributions] Experiments were 
 performed primarily in Pittsburgh by T.\ Tuan,
  and in Bordeaux by H.\ Kellay.
 Analysis of data was carried out by all authors.  
 Research was designed by G.\ Gioia, N.\ Goldenfeld, and W.\ Goldburg.
  G.\ Gioia wrote the paper with assistance 
 from N.\ Goldenfeld and P.\ Chakraborty 
 (who also wrote the Supplementary Information). 
\end{addendum}

\end{document}